\documentclass{mnras}
\usepackage{graphicx}
\usepackage{txfonts}
\let\bibhang\relax
\usepackage{natbib}
\bibpunct{(}{)}{;}{a}{}{,}

\title[HD~76582's circumstellar disk]{Far-infrared and sub-millimetre imaging of HD~76582's circumstellar disk}
\author[J.~P. Marshall et al.]{Jonathan P. Marshall$^{1,2}$\thanks{E-mail: jonty.marshall@unsw.edu.au}, Mark Booth$^{3}$, Wayne Holland$^{4,5}$, Brenda C. Matthews$^{6,7}$, \newauthor Jane S. Greaves$^{8}$, Ben Zuckerman$^{9}$\\
$^{1}$School of Physics, UNSW Australia, High Street, Kensington, Sydney, NSW 2052, Australia\\
$^{2}$Australian Centre for Astrobiology, UNSW Australia, High Street, Kensington, Sydney, NSW 2052, Australia\\
$^{3}$Instituto de Astrof\'isica, Pontificia Universidad Cat\'olica de Chile, Vicua Mackenna 4860, 7820436 Macul, Santiago, Chile\\
$^{4}$UK Astronomy Technology Center, Royal Observatory, Blackford Hill, Edinburgh EH9 3HJ, UK\\
$^{5}$Institute for Astronomy, University of Edinburgh, Royal Observatory, Blackford Hill, Edinburgh EH9 3HJ, UK\\
$^{6}$National Research Council of Canada, 5071 West Saanich Road, Victoria, BC V9E 2E7, Canada\\
$^{7}$University of Victoria, Finnerty Road, Victoria, BC V8W 3P6, Canada\\
$^{8}$School of Physics and Astronomy, Cardiff University, Queen's Buildings, The Parade, Cardiff, CF24 3AA, UK\\
$^{9}$Department of Physics and Astronomy, University of California, Los Angeles, CA 90095, USA}
\begin{document}

\date{Accepted ---. Received ---; in original form ---}

\pagerange{000--000} \pubyear{---}

\maketitle

\label{firstpage}

\begin{abstract}
Debris disks, the tenuous rocky and icy remnants of planet formation, are believed to be evidence for planetary systems around other stars. The JCMT/SCUBA-2 debris disk legacy survey `SCUBA-2 Observations of Nearby Stars' (SONS) observed 100 nearby stars, amongst them HD~76582, for evidence of such material. Here we present imaging observations by JCMT/SCUBA-2 and \textit{Herschel}/PACS at sub-millimetre and far-infrared wavelengths, respectively. We simultaneously model the ensemble of photometric and imaging data, spanning optical to sub-millimetre wavelengths, in a self-consistent manner. At far-infrared wavelengths, we find extended emission from the circumstellar disk providing a strong constraint on the dust spatial location in the outer system, although the angular resolution is too poor to constrain the interior of the system. In the sub-millimetre, photometry at 450 and 850~$\mu$m reveal a steep fall-off that we interpret as a disk dominated by moderately-sized dust grains ($a_{\rm min}~=~36~\mu$m), perhaps indicative of a non-steady-state collisional cascade within the disk. A disk architecture of three distinct annuli, comprising an unresolved component at ~ 20 au and outer components at 80 and 270 au, along with a very steep particle size distribution ($\gamma~=~5$), is proposed to match the observations. 
\end{abstract}


\section{Introduction}

The presence of a circumstellar dust disk around a mature, main-sequence star is interpreted as a visible signpost of a planetary system \citep{Matthews2014}. These discs are the remnants of asteroidal and cometary bodies broken up in mutual collisions and so are commonly known as `debris discs' \citep{BackPar1993,Krivov2010}. Debris discs are comprised of icy and rocky bodies ranging from sub-micron sized dust grains to kilometre sized planetesimals, analogous to the solar system's Asteroid belt and Edgeworth-Kuiper belt \citep{Wyatt2008}. Between 20 and 30 per cent of nearby, main-sequence AFGK-type stars have been identified as hosting cool debris disks, with fractional luminosities ($L_{\rm IR}/L_{\star}$) ranging from $10^{-3}$ to $10^{-6}$ \citep{Eiroa2013,Thureau2014}. The observed disk incidence is sensitivity limited, with current instrumentation unable to directly detect a disk of equivalent brightness to the Edgeworth-Kuiper belt around another star \citep[$L_{\rm IR}/L_{\star} \sim 1.2\times10^{-7}$][]{Vitense2012}. Understanding these structures is fundamental to obtaining a full picture of the outcomes of planet formation processes, and the evolution of planetary systems \citep[e.g.][]{Marshall2014a,MoroMar2015,WitMar2015}.

The debris disk around HD~76582 was identified by detection of infrared excess from the star at 60~$\mu$m in the all-sky survey by the \textit{InfraRed Astronomical Satellite} \citep[\textit{IRAS};][]{Neugebauer1984}. Its fractional luminosity of $1.7~\pm~0.2~\times10^{-4}$, estimated from a single temperature blackbody fit with a temperature of 80~K \citep{ZuckSong2004,Moor2006}, is consistent with a steady-state collisional cascade according to the evolutionary models of \cite{DomDec2003} assuming an age of $\sim$ 0.5~Gyr \citep{Moor2006,Rhee2007,David2015}. However, if the age of $\sim$ 2~Gyr from \cite{ZorecRoyer2012} is considered, then the disk appears to be unusually bright. 

Images at longer wavelengths trace larger and colder dust grains, making them useful in identifying the location of the dust producing planetesimal belts around a disk-host star \citep{Krivov2008}. Additionally, (sub-)millimetre wavelength observations are ideal for probing asymmetric structures formed by resonance-trapped dust grains in disks, possibly caused by perturbing planets \citep[e.g.][]{Wyatt2003,Hughes2011,LesThil2015}. In the case of HD~76582, the data presented here are low signal-to-noise (5-10-$\sigma$ integrated detection) and the disk is unresolved. The sub-millimetre data are critical to the disk modelling process, constraining the dust particle size distribution from the SED's sub-millimetre slope for the first time, and providing a measure of the dust mass in the system.

The \textit{Spitzer}/IRS spectrum, spanning mid-infrared wavelengths, reveals the presence of a second, warm component to this disk \citep{Chen2014}. Debris disks with spectral energy distributions (SEDs) exhibiting two distinct temperatures are inferred to comprise two physically distinct planetesimal rings that produce the dust at different temperatures \citep{Morales2011}. A favoured mechanism invoked for keeping the space between the two planetesimal rings free of migrating dust is the presence of one (or more) planet(s) \citep{SuRie2013}. Determination of the architectures of these systems solely from their SEDs is limited by the contrast between the warm and cool dust components \citep{Kennedy2014}. Such interpretation is crude, and even combining the SED with moderately resolved disk images cannot always provide a clear architecture from modelling \citep[e.g.][]{Ertel2014}. High angular resolution images are therefore critical to interpretation of these systems.

In this work we combine these disparate data to produce a self-consistent and coherent model of the disk's architecture and evolutionary state. In Section 2 we present the observations used to model this system, including the new SCUBA-2 sub-millimetre images and ancillary data from the literature. In Section 3 we present a summary of the simultaneous modelling of imaging and photometric data. In Section 4 we discuss the outcomes of the modelling process, interpreting the properties of HD~76582's disk as part of the ensemble of known debris disks. Finally in Section 5 we present our conclusions.

\section{Observations}

Here we summarise the stellar physical parameters and the available photometric measurements of the HD~76582 system. These were taken from a variety of literature sources and combined with new data from the \textit{Herschel}/PACS and JCMT/SCUBA-2 observations in the modelling process. 

\subsection{Stellar parameters and photosphere}

HD~76582 (HIP~44001, $o^{2}$ Cnc) is an F0~IV type star located at a distance of 46.13~$\pm$~0.69~pc. The distance was taken from the re-reduction of \textit{HIPPARCOS} data by \cite{vanLeeuwen2007}. The spectral type is taken from \cite{Skiff2014}. Studies have found no evidence of binarity or multiplicity for the system \citep{EggTok2008,DeRosa2014}.  The absolute magnitude and bolometric correction are taken from \cite{AvE2012}. The effective temperature, $T_{\rm eff}$, surface gravity, $\log g$, stellar radius, $R_{\star}$, and stellar mass, $M_{\star}$, are taken from \cite{AP1999}. The metallicity, [Fe/H], is taken from \cite{Giridhar2013}. The rotational velocity $\nu~\sin i$ and the rotational period (modulo the unknown stellar inclination) $P_{\rm rot}/\sin i$ are taken from \cite{AvE2012}. Stellar age estimates in the literature range from 300~Myr, based on $UVW$ space velocities \citep{ZucSong2004} and HR diagram location \citep{Lowrance2000}, or 450$^{+150}_{-290}$~Myr \citep{Moor2006}, to 2.13~$\pm$~0.24~Gyr \citep{ZorecRoyer2012}. The stellar physical parameters used in this work are summarised in Table \ref{table:star_parameters}. 

\begin{table}
\caption{Stellar physical properties. \label{table:star_parameters}}
\centering
\begin{tabular}{lr}
\hline\hline
Parameter & HD~76582 \\
\hline
Distance [pc] & 46.13~$\pm$~0.69 \\
Right Ascension$^{a}$ [h:m:s] & 08$^{\rm h}$57$^{\rm m}$35\fs200 \\
Declination$^{a}$ [d:m:s] & +15\degr34\arcmin52\farcs61 \\
Spectral Type & F0~IV \\
V, B-V [mag] & 5.687, 0.207 \\
$M_{V}$, B.C.$^{b}$ [mag] & 2.580, -0.03 \\
$L_{\star}$ [$L_{\odot}$] & 10.30~$\pm$~0.43 \\
$T_{\rm eff}$ [K] & 7868 \\
$\log g$ [cm/s$^{2}$] & 4.25 \\
$R_{\star}$ [$R_{\odot}$] & 1.62~$\pm$~0.08 \\
$M_{\star}$ [$M_{\odot}$] & 1.72~$\pm$~0.01 \\
$[\rm{Fe/H}]$ & 0.2 \\
$\nu \sin i$ [km/s] & 90.5~$\pm$~4.5 \\
$P_{\rm rot}/\sin i$ [days] & 0.8 \\
Age [Gyr] & 0.30 to 2.13 \\
\hline
\end{tabular}

\medskip
\raggedright
\textbf{Notes:} (a) Coordinates are in ICRS J2000 epoch (b) Bolometric Correction.
\end{table}

A Kurucz stellar photosphere model \citep{CasKur2004} was scaled to the optical \citep{HauckMerm1998,Perryman1997}, near- and mid-infrared \citep{Skrutskie2006,Egan2003,Wright2010} measurements at wavelengths $<$~15~$\mu$m (where no clear excess is visible) using a least-squares fit to the measured flux densities, weighted by their individual uncertainties. The best fitting model was subsequently extrapolated into the far-infrared and sub-millimetre in order to quantify the stellar contribution to the total emission at those longer wavelengths.

\subsection{Photometry and Spectroscopy}

\begin{table}
\caption{Photometry used in modelling. \label{table:photometry}}
\centering
\begin{tabular}{lclc}
\hline\hline
\multicolumn{1}{c}{Wavelength} & Flux Density & Telescope \& & Reference \\
\multicolumn{1}{c}{$[\mu \rm{m}]$}  &    [mJy]     & Instrument &           \\
\hline
\phantom{00}0.349 & \phantom{0}5555~$\pm$~\phantom{0}15 & Str\"omgren $u$ & 1 \\
\phantom{00}0.411 & 17290~$\pm$~\phantom{0}64  & Str\"omgren $v$ & 1 \\
\phantom{00}0.440 & 17710~$\pm$~\phantom{0}81  & Johnson $B$ & 2 \\
\phantom{00}0.466 & 20340~$\pm$~\phantom{0}75  & Str\"omgren $b$ & 1 \\
\phantom{00}0.546 & 19720~$\pm$~181 & Str\"omgren $y$ & 1 \\
\phantom{00}0.550 & 19180~$\pm$~179 & Johnson $V$ & 2 \\
\phantom{00}0.790 & 16070~$\pm$~147 & Cousins $I$ & 2 \\
\phantom{00}1.235 & 12430~$\pm$~261 & 2MASS $J$ & 3 \\
\phantom{00}1.662 &  \phantom{0}8455~$\pm$~139 & 2MASS $H$ & 3 \\
\phantom{00}2.159 &  \phantom{0}5623~$\pm$~123 & 2MASS $K_{\rm S}$& 3 \\
\phantom{00}4.29  &  \phantom{0}1538~$\pm$~149 & \textit{MSX} B1 & 4 \\
\phantom{00}4.35  &  \phantom{0}1496~$\pm$~156 & \textit{MSX} B2 & 4 \\
\phantom{00}4.60  &  \phantom{0}1742~$\pm$~130 & \textit{WISE} W2 & 5 \\
\phantom{00}8.28  &  \phantom{00}463~$\pm$~\phantom{0}24  & \textit{MSX} A & 4 \\
\phantom{00}9.0   &  \phantom{00}392~$\pm$~\phantom{0}30  & \textit{AKARI}/IRC & 6 \\
\phantom{0}11.6   &  \phantom{00}262~$\pm$~\phantom{00}4 & \textit{WISE} W3 & 5 \\
\phantom{0}12.13  &  \phantom{00}217~$\pm$~\phantom{0}15 & \textit{MSX} C & 4 \\
\phantom{0}13.0   &  \phantom{00}201~$\pm$~\phantom{0}14 & \textit{Spitzer}/IRS & 7 \\
\phantom{0}14.65  &  \phantom{00}151~$\pm$~\phantom{0}13 & \textit{MSX} D & 4 \\
\phantom{0}18.0   &  \phantom{00}121~$\pm$~\phantom{0}12 & \textit{Spitzer}/IRS & 7 \\
\phantom{0}22.1   &  \phantom{000}93~$\pm$~\phantom{00}3 & \textit{WISE} W4 & 5 \\
\phantom{0}24.0   &  \phantom{000}87~$\pm$~\phantom{00}9 & \textit{Spitzer}/IRS & 7 \\
\phantom{0}31.0   &  \phantom{000}87~$\pm$~\phantom{0}15 & \textit{Spitzer}/IRS & 7 \\
\phantom{0}60.0   &  \phantom{00}391~$\pm$~\phantom{0}47 & \textit{IRAS} 60 & 8 \\
100.0 & \phantom{00}609~$\pm$~\phantom{0}43 & \textit{Herschel}/PACS & 7 \\
160.0 & \phantom{00}495~$\pm$~\phantom{0}35 & \textit{Herschel}/PACS & 7 \\
450.0 & \phantom{00}107~$\pm$~\phantom{0}25 & JCMT/SCUBA-2 & 7 \\
850.0 & \phantom{00}5.7~$\pm$~1.0 & JCMT/SCUBA-2 & 7 \\
\hline 
\end{tabular}

\medskip
\raggedright
\textbf{References:} (1) \cite{HauckMerm1998}; (2) \cite{Perryman1997}; (3) \cite{Skrutskie2006}; (4) \cite{Egan2003}; (5) \cite{Wright2010}; (6) \cite{Ishihara2010}; (7) This work; (8) \cite{Beichman1988}.
\end{table}

In addtion to the sub-millimetre images and photometry from JCMT/SCUBA-2, we compiled literature photometry spanning the mid- and far-infrared to characterise the debris disk's thermal emission. Of particular note are the mid-infrared \textit{Spitzer}/IRS spectrum and far-infrared \textit{Herschel}/PACS images, both of which provide complementary data, namely tracing the rise of the disk emission above the stellar photosphere, and the peak emission and spatial extent of the disk, respectively. These measurements are critical to obtaining a comprehensive overview of the disk architecture. A summary of the photometry used in modelling the star and debris disk is presented in Table \ref{table:photometry}.

\subsubsection{Mid-infrared}

We use the \textit{MSX} \citep{Egan2003}, \textit{AKARI} \citep{Ishihara2010}, and \textit{WISE} \citep{Wright2010} photometry to constrain the mid-infrared SED. A \textit{Spitzer} \citep{Werner2004} InfraRed Spectrograph \citep[IRS;][]{Houck2004} observation of HD~76582 (Program ID 40651, P.I. Houck) spanning 5 to 38~$\mu$m was obtained from the CASSIS archive\footnote{The Cornell Atlas of Spitzer/IRS Sources (CASSIS) is a product of the Infrared Science Center at Cornell University, supported by NASA and JPL.} \citep{Lebouteiller2011}. The spectrum was scaled to the stellar photosphere model between 8 and 12~$\mu$m by least-squares fitting. A factor of 1.04 (increase of 4 per cent) was applied to the IRS spectrum in order to match the stellar photosphere model. This is in-line with typical scaling applied to calibrated IRS spectra \citep[2\% to 7\%,][]{Chen2014}. 

For inclusion in the SED fitting, flux densities were measured along the spectrum at wavelengths of 9, 13, 18, 24 and 31~$\mu$m. For each measurement the spectrum was averaged over all data points within 2~$\mu$m of the nominal wavelength. The IRS spectrum thus shows evidence for mid-infrared excess, strongly rising from the stellar photosphere at wavelengths beyond 20~$\mu$m.

\subsubsection{Far-infrared}

\begin{figure*}
\centering
\includegraphics[width=0.8\textwidth]{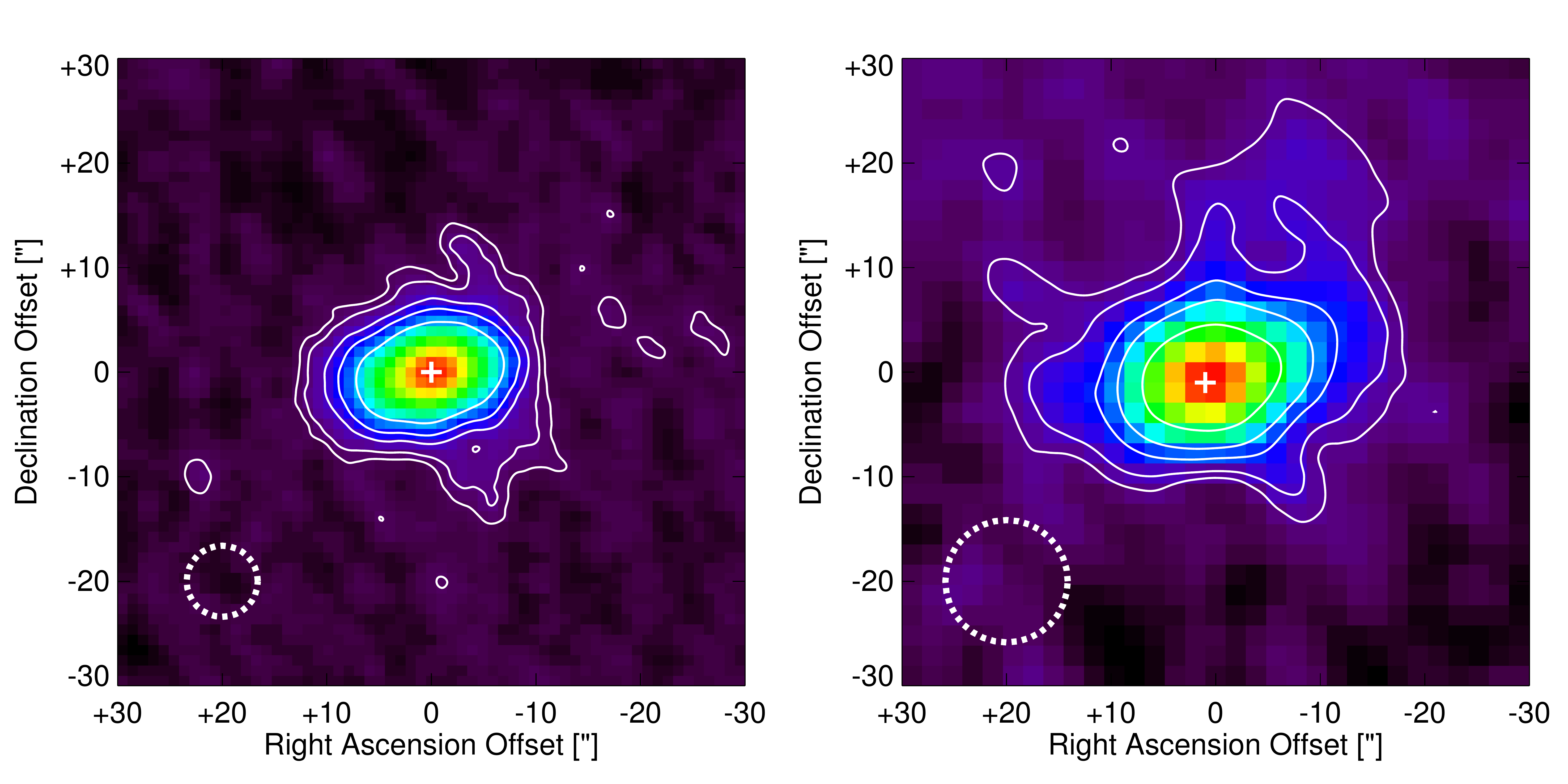}
\caption{\textit{Herschel}/PACS images of HD~76582 at 100~$\mu$m (left) and 160~$\mu$m (right). Stellar position is denoted by the white `+' symbol. Contours denote 3, 5, 10, 15, and 25-$\sigma$ steps for a background r.m.s. of 0.07~mJy/arcsec$^{2}$ at 100~$\mu$m and 0.15~mJy/arcsec$^{2}$ at 160~$\mu$m. Image scale is 1\arcsec~per pixel at 100~$\mu$m and 2\arcsec~per pixel at 160~$\mu$m. Instrument beam FWHM is denoted by the dashed circle in the bottom left corner. Orientation is north up, east left. \label{figure:pacs_images}}
\end{figure*}

\textit{Herschel}\footnote{\textit{Herschel} is an ESA space observatory with science instruments provided by European-led Principal Investigator consortia and with important participation from NASA.} \citep{Pilbratt2010} Photometer Array Camera and Spectrograph \citep[PACS;][]{Poglitsch2010} scan-map observations of HD~76582 at 100 and 160~$\mu$m (Programme ot2\_bzuckerm\_2, P.I. Zuckerman) were obtained from the \textit{Herschel} Science Archive. The \textit{Herschel}/PACS images are presented in Fig. \ref{figure:pacs_images}.

The individual scans were reduced interactively from the level 0 (raw data) products and mosaicked using the \textit{Herschel} Interactive Processing Environment\footnote{HIPE is a joint development by the \textit{Herschel} Science Ground Segment Consortium, consisting of ESA, the NASA \textit{Herschel} Science Center, and the HIFI, PACS and SPIRE consortia.} \citep[HIPE;][]{Ott2010} version 12, and PACS calibration version 65. We followed the reduction method adopted for the \textit{Herschel} Open Time Key Programme `Dust around Nearby Stars'; see \cite{Eiroa2013} for a detailed explanation. The data were high-pass filtered, with widths of 82\arcsec~at 100~$\mu$m and 102\arcsec~at 160~$\mu$m, to suppress large scale background structure in the final images. Image scales of 1\arcsec~per pixel at 100~$\mu$m and 2\arcsec~per pixel at 160~$\mu$m were used in the final mosaics. 

Flux densities were measured using aperture photometry, with appropriate aperture corrections applied to the measurement \citep[taken from][]{Balog2014} and background estimation via multiple sky apertures, again following \cite{Eiroa2013}. We obtain flux density measurements of 609~$\pm$~43~mJy at 100~$\mu$m, and 495~$\pm$~35~mJy at 160~$\mu$m.

\subsubsection{Sub-millimetre}

\begin{figure*}
\centering
\includegraphics[width=0.8\textwidth]{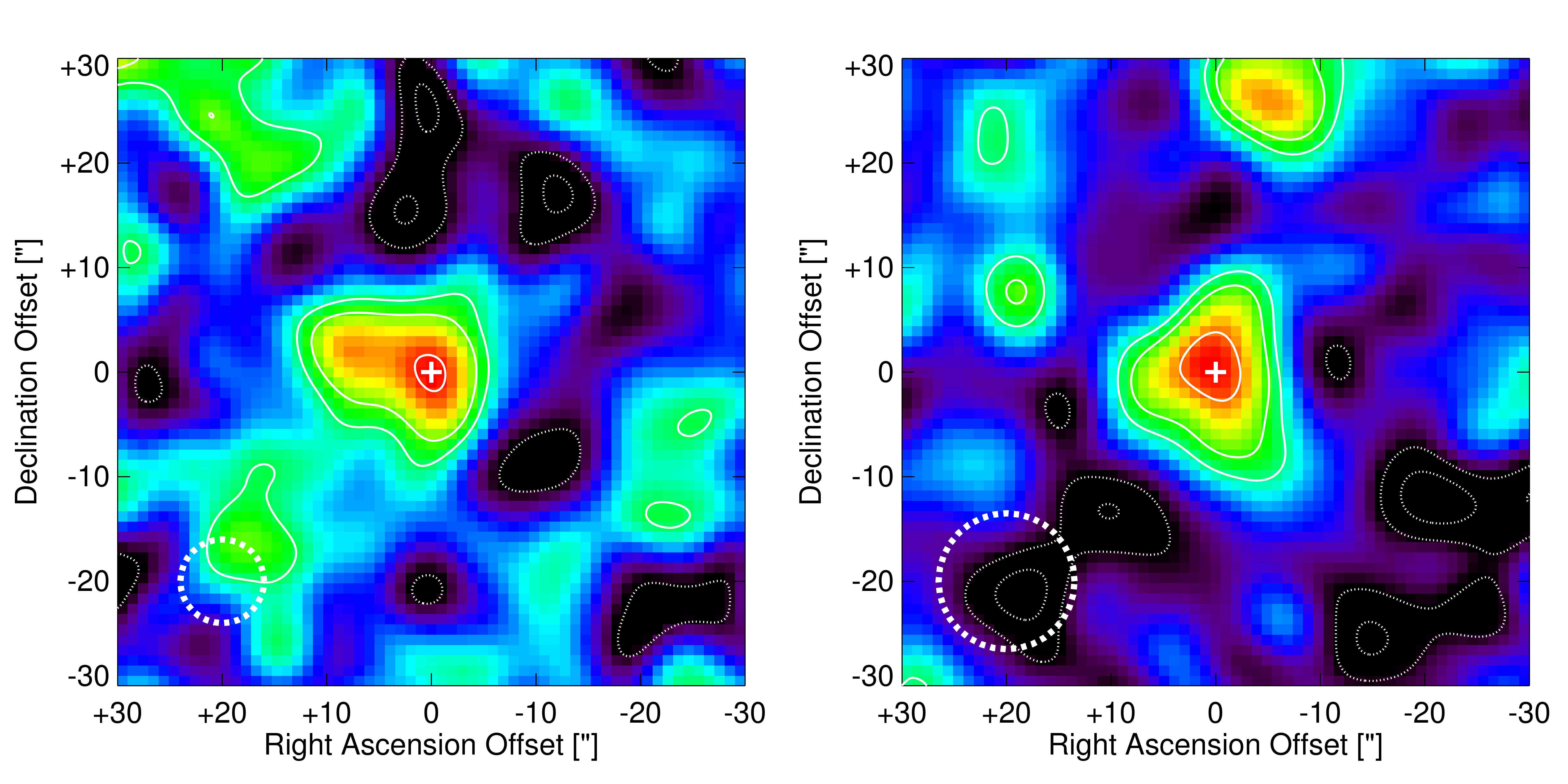}
\caption{JCMT/SCUBA-2 images of HD~76582 at 450~$\mu$m (left) and 850~$\mu$m (right). The images have been smoothed with a 7\arcsec~FWHM Gaussian. Stellar position is denoted by the white `+' symbol. Contours denote -3, -2, 2, 3, and 5-$\sigma$ steps for a background r.m.s. of 0.306~mJy/arcsec$^{2}$ at 450~$\mu$m and 0.005~mJy/arcsec$^{2}$ at 850~$\mu$m. Image scale is 1\arcsec~per pixel. Instrument beam is denoted by the dashed circle in the bottom left corner. Orientation is north up, east left. \label{figure:scuba-2_images}}
\end{figure*}

The ames Clerk Maxwell Telescope's Sub-millimetre Common Use Bolometer Array instrument \citep[JCMT/SCUBA-2;][]{Holland2013} 450 and 850~$\mu$m data for HD~76582 consist of five hours of observations taken in ten half-hour chunks. The observations were taken using the constant speed DAISY pattern, providing uniform exposure time coverage of a region $\sim$~3\arcmin~in diameter centred on the target. The SCUBA-2 observations were reduced using the Dynamic Interative Map-Maker within the STARLINK SMURF package \citep{Chapin2013} called from the ORAC-DR automated pipeline \citep{Cavanagh2008}. The map maker configuration file was optimised for compact sources with a known position. A `zero-masking' technique was applied to constrain the mean value of the map to zero beyond 60\arcsec~from the centre of the field for all but the final iteration of the map-making process. This aided convergence of the iterative map-making and suppressed large-scale ringing artefacts. The data were high-pass filtered at 1~Hz ($\equiv$~150\arcsec), removing large spatial scale noise. A final mosaic was created at each wavelength by co-adding the individual maps using inverse variance weighting. The mosaics have a scale of 1\arcsec~per pixel. The mosaics have been smoothed with a 7\arcsec~FWHM Gaussian to improve the signal-to-noise. The images are presented in Fig. \ref{figure:scuba-2_images}.

In the 450~$\mu$m image there is a clear peak, well-centered on the stellar position. The peak flux density is 89~$\pm$~17~mJy/beam. Although the disk looks extended this is not significant. Negative `holes' are visible in the map close to the target position, and complicate aperture photometry measurement (see Fig. \ref{figure:scuba-2_images}). For a 12.5\arcsec~radius aperture we obtain a flux density of 107~$\pm$~25~mJy, consistent with the peak flux density measurement. At 850~$\mu$m there is a clear peak $\sim$3\arcsec~offset from the stellar position. Such an offset is consistent with the magnitude of shifts expected for low signal-to-noise data, and much smaller than the beam FWHM (13\arcsec). The peak flux density is 5.8~$\pm$~1.0~mJy/beam.

The observations of HD~76582 were obtained in two observing runs. From the first chunk of two hours of data, the disk is visible at 450~$\mu$m (188~$\pm$~44~mJy), but not at 850~$\mu$m (3-$\sigma~<~4.8~$mJy). Such a result is unusual, as the greater sensitivity of JCMT/SCUBA-2 at 850~$\mu$m typically results in non-detection of debris disks at 450~$\mu$m. Furthermore, the slope of the sub-millimetre SED implied by this non-detection was far greater than the values obtained from previous detections or expected from theoretical considerations. Additional observations totalling three hours were acquired to determine the exact nature of the sub-millimetre excess, which resulted in the disk's detection at 850~$\mu$m, but also confirmed the steepness of the sub-millimetre slope, albeit less extreme than was measured in the intial observations.

\section{Source brightness profiles}

\begin{figure*}
\centering
\includegraphics[width=0.8\textwidth]{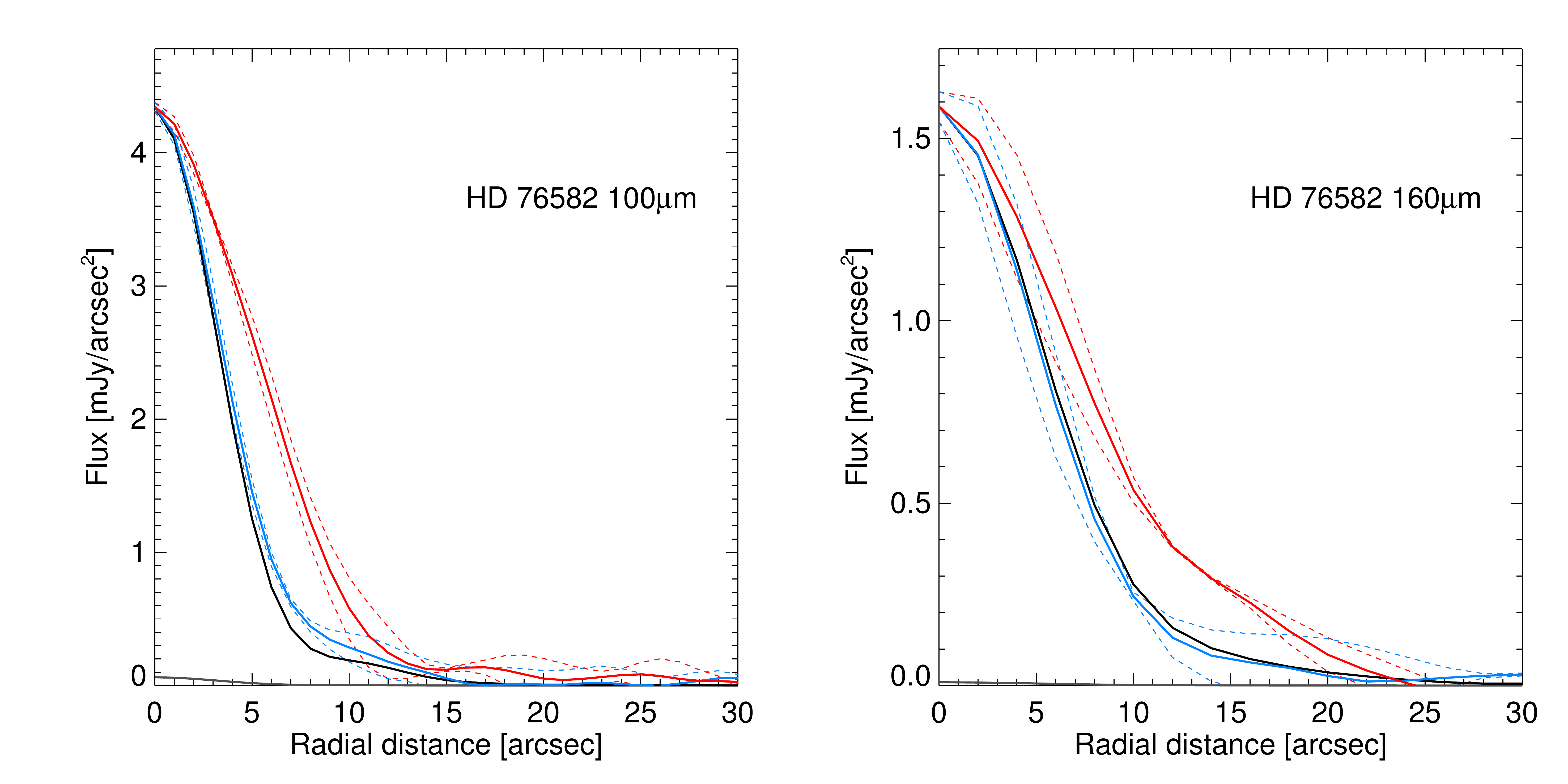}
\caption{Radial profiles of HD~76582 at 100~$\mu$m (left) and 160~$\mu$m (right). The solid red line denotes the disk major axis, whilst the blue solid line denotes the minor axis; dashed lines of each colour represent the upper and lower bounds of the 1-$\sigma$ uncertainties. The grey line denotes a PSF scaled to the stellar photospheric contribution. The black line is a PSF scaled to the peak disk emission. \label{figure:pacs_radprofs}}
\end{figure*}

We measured the source brightness profile along the major and minor axes in both \textit{Herschel} bands as an additional constraint for the disk modelling process, see Fig. \ref{figure:pacs_radprofs}. To measure the profiles, the image was first rotated by an angle such that its major axis lay along the image $x$-axis. This rotated image was then interpolated to a grid 10$\times$ finer than the image pixel scale (0.1\arcsec/0.2\arcsec per element at 100~$\mu$m/160~$\mu$m). The radial profile was measured from the peak of the source brightness profile to an extent of 30\arcsec~at intervals equivalent to the pixel scale of the image (1\arcsec/2\arcsec~at 100~$\mu$m/160~$\mu$m). The mean and standard deviation of two areas equidistant from the source peak along the $x$/$y$-axis were used to calculate the major/minor axis profile and its uncertainty, respectively. 

To determine the disk extent, we follow the procedure laid out in \cite{Marshall2014b}. A two dimensional Gaussian is fitted to the source brightness profile in each image to estimate the source FWHM along the major and minor axes, $A_{\rm image}$ and $B_{\rm image}$, respectively, and the disk position angle, $\theta$. The disk is then deconvolved from the instrument beam using a PSF standard, in this case an image of $\alpha$ B\"ootis, reduced in the same way as the images and rotated to the same orientation angle as the observations. The radially averaged FWHMs of the PSF are 6.8\arcsec~at 100~$\mu$m and 11.3\arcsec~at 160~$\mu$m. The deconvolved major and minor axis of the disk, $A_{\rm disk}$ and $B_{\rm disk}$ are then measured, again using a two dimensional Gaussian profile. The disk is broader than the PSF along both axes in the PACS 100~$\mu$m images and the disk inclination is obtained from the deconvolved extent, i.e. $i\!=\!\cos^{-1}(B_{\rm disk}/A_{\rm disk})$. However in the PACS 160~$\mu$m images, the disk is only broader along the major axis and thus only marginally resolved in that waveband.  A summary of the disk geometry and orientation derived from this analysis is presented in Table \ref{table:disc_props}. The (minor) differences between the extent of the disk obtained here and in Vican et al. (e.g. lack of extended emission along the minor axis at 160~$\mu$m) could be due to the choice of $\alpha$ Boo rather than $\alpha$ Cet as a PSF standard.

\begin{table}
\centering
\caption{Orientation and extent (FWHM or diameter) of the disk. \label{table:disc_props}}
\begin{tabular}{lcc}
\hline\hline
 & \multicolumn{2}{c}{HD~76582} \\
Parameter & 100~$\mu$m & 160~$\mu$m \\
\hline
$A_{\rm image}$ [\arcsec] & 11.8~$\pm$~0.5 & 16.2~$\pm$~0.7 \\
$B_{\rm image}$ [\arcsec] & \phantom{0}8.0~$\pm$~0.4 & 11.2~$\pm$~0.5 \\
PSF [\arcsec] & 6.8 & 11.3 \\
$A_{\rm disk}$ [\arcsec] & 5.9~$\pm$~0.1 & 8.5~$\pm$~0.1 \\ 
$A_{\rm disk}$ [au] & 271~$\pm$~10 & 391~$\pm$~14 \\
$\theta$ [\degr] & 103~$\pm$~\phantom{0}5 & 115~$\pm$~\phantom{0}5 \\
$i$ [\degr] & \multicolumn{2}{c}{64~$\pm$~4} \\
\hline
\end{tabular}
\end{table}

\section{Modelling}

In the first instance we use a modified blackbody model to reproduce the observed excess from the circumstellar disk. Although not physcially meaningful per se, this approach provides some guidance as to the number of debris components present in the system, and their location(s). We then apply those findings to constrain the parameter space explored by a power law disk model that uses both the SED and images (radial profiles) in the fitting process. 

\subsection{Modified blackbody}

The disk temperature derived from a blackbody fit to photometry up to 60~$\mu$m predicts a (minimum) dust semi-major axis of 43~au from the star, although more recent work by \citet{Pawellek2014} on resolved disks suggests the actual extent should be about twice this for a star of $\sim$ 10~$L_{\odot}$. The \textit{Herschel}/PACS far-infrared images of HD~76582 at 100 and 160~$\mu$m presented here and in Vican et al. (submitted) show emission from a broad disk extending several hundred au from the star. 

We attempt to replicate the shape of the disk SED with both one and two component modified blackbody models. The model is defined by a Planck function, $B(\nu,T_{\rm dust})$, for each component and a critical wavelength, $\lambda_{\rm 0}$, beyond which the Planck function is modified by a factor $(\lambda/\lambda_{\rm 0})^{-\beta}$ reducing the flux density at wavelengths greater than $\lambda_{\rm 0}$. In both cases, the model is scaled to match the observed flux density at 100~$\mu$m. 

\begin{figure*}
\centering
\includegraphics[trim={0 .5cm 0 .5cm},clip,width=0.8\textwidth]{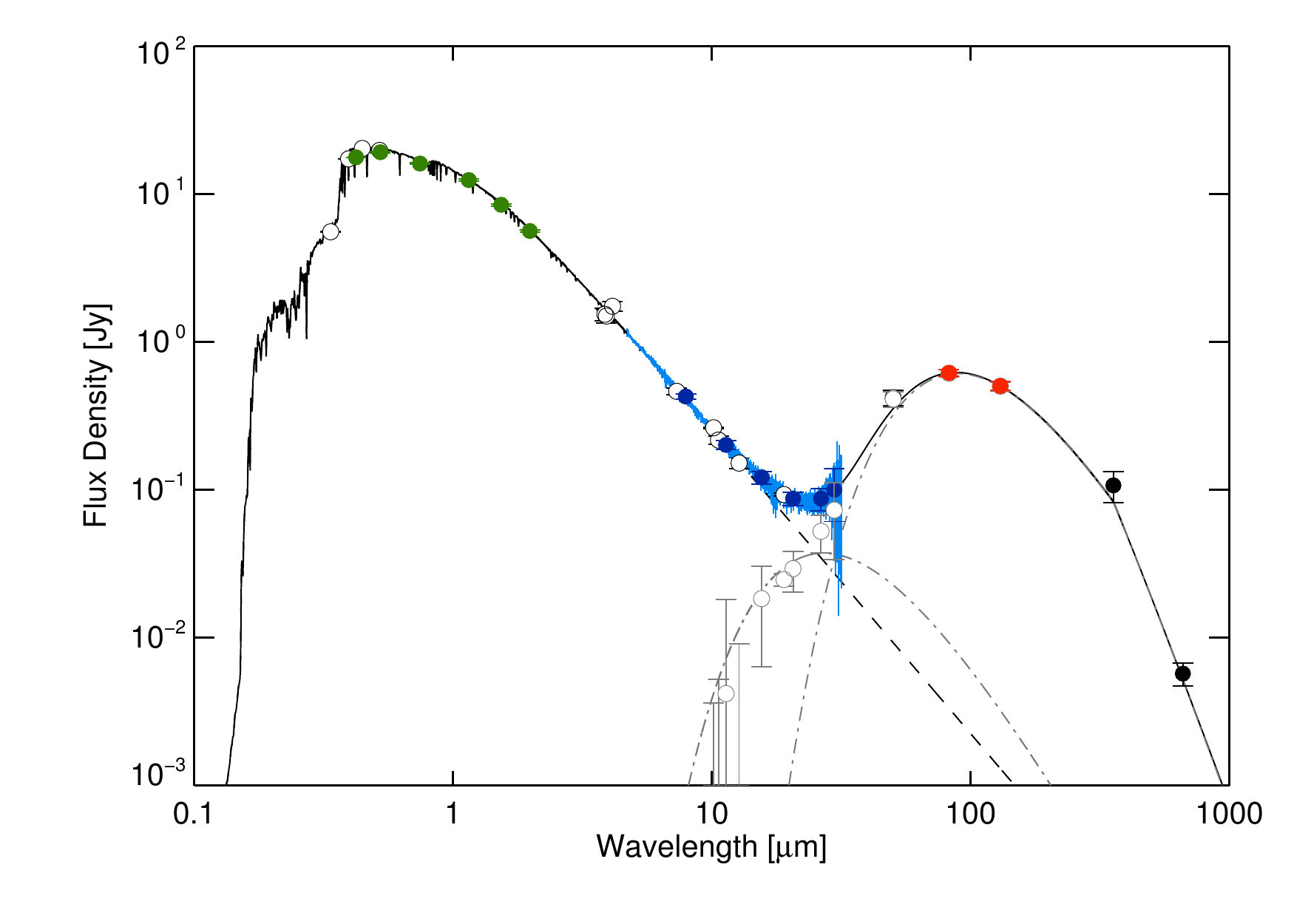}
\caption{Two component modified blackbody SED fit. Green denotes data used to scale the model stellar photosphere, light and dark blue denote the \textit{Spitzer}/IRS spectrum and photometry respectively, red denotes \textit{Herschel}/PACS photometry, black denotes JCMT/SCUBA-2 photometry, white denotes literature data, and grey denotes photosphere subtracted photometry. The dashed line is the stellar photosphere, the dot-dash lines are the disk components, and the solid line is the combined star+disk model. \label{figure:sed_two_part_modbb}}
\end{figure*}

A single component modified blackbody cannot simultaneously replicate both the rise of the disk from the photosphere at mid-infrared wavelengths, and the far-infrared excess/sub-millimetre slope. A two component model replicates the observed excess satisfactorially, although not perfectly, across all wavelengths. This is indicative of either the presence of a single broad disk around HD~76582, or multiple physical components to the disk. The parameters of the best fitting model are presented in Table \ref{table:mbb_results}, whilst the SED itself is shown in Fig. \ref{figure:sed_two_part_modbb}. 

\begin{table}
\centering
\caption{Results of modified blackbody fit to SED. \label{table:mbb_results}}
\begin{tabular}{lccc}
\hline\hline
Parameter & Range & Distribution & Best Fit \\
\hline
$T_{\rm inner}$ [K] & \phantom{0}60 -- 260 & Linear & 150~$\pm$~20 \\
$T_{\rm outer}$ [K] & \phantom{0}30 -- \phantom{0}60 & Linear & 44~$\pm$~5 \\
$\lambda_{\rm 0}$ [$\mu$m] & 450 & Fixed & 450 \\
$\beta$ & \phantom{00}0 -- \phantom{00}2 & Linear & 1.9~$\pm$~0.5 \\
$L_{\rm IR}/L_{\star}$ [$\times10^{-4}$] & n/a & Continuous & 1.9~$\pm$~0.3 \\
$M_{\rm dust}$ [$\times10^{-2}~M_{\oplus}$] & n/a & Continuous & 8.4~$\pm$~1.4 \\
$\chi^{2}_{\rm red}$ & n/a & n/a & 0.97 \\
\hline
\end{tabular}
\end{table}

A simultaneous least-squares fit of two blackbodies to the excess at $\lambda~\geq~20~\mu$m and s/n$~\geq~$3 was used to estimate the temperatures of the disk components. We find best-fit temperatures of 150~$\pm$~20~K for the inner component, and 44~$\pm$~5~K for the outer component. In the blackbody approximation, these translate to separations of around 10 and 130~au, respectively. The fractional luminosities of the inner and outer components were 3$\times10^{-5}$ and 1.6$\times10^{-4}$. Uncertainties were estimated from the shape of the $\chi^{2}$ distribution by varying parameters individually whilst keeping the remainder fixed at the values of the best-fit model. For the outer radius, this is close to the disk major axis derived from the extended emission in the \textit{Herschel} images, $A_{\rm disk} = 271~\pm~10~$au.

Due to the sparse coverage at sub-millimetre wavelengths, the break wavelength, $\lambda_{\rm 0}$, is poorly constrained by the modelling. We therefore fix $\lambda_{\rm 0}$ to 450~$\mu$m. Typically, $\lambda_{\rm 0}$ is taken as 210~$\mu$m; our adoption of a longer wavelength favours models with a warmer outer component leading to a larger $\beta$ parameter to match the SED's slope between 450 and 850~$\mu$m. The $\beta$ parameter itself, lying $\sim$~2, is consistent with typical values for small dust grains in the interstellar medium, rather than those in circumstellar disks which would be closer to $\beta\!\sim\!1$. A steep sub-millimetre slope usually signifies a disk whose SED is dominated by the smaller grains. In order to tally this with the agreement found between the imaged disk extent and the blackbody approximation from the SED we might invoke a large minimum grain size for the dust; this will be dealt with in the next section. A total disk fractional luminosity of 2.3$\times10^{-4}$ is calculated from the sum of the two components, slightly higher than, but comparable to, previous estimates \citep{Moor2006}.

An estimate for the dust mass in the disk can be obtained from the 850~$\mu$m flux density using the relation $M_{\rm dust}\!=\!F_{850}~d^{2} / \kappa_{850}~B_{\nu,T_{\rm dust}}$ \citep[e.g.][]{ZucBec1993,NajWil2005,Panic2013}. Assuming a dust opacity of 1.7~gcm$^{-2}$ \citep[e.g.,][]{Pollack1994}, a dust mass of 8.4~$\pm$~1.4~$\times10^{-2}~M_{\oplus}$ is obtained for HD~76582's disk. This estimate is subject to large uncertainties due to the unknown dust grain properties such as emissivity, structure and size distribution, such that the dust mass derived here may be a factor of 3 to 5 lower than the actual value \citep{NajWil2005}.

\subsection{Power law}

Here we approximate the disk with a power law surface density (and size distribution), astronomical silicate dust model. We use the radiative transfer code SEDUCE \citep{Mueller2010} to fit the SED. Model disk images were created based on the input parameters to the SED and convolved with an instrumental PSF for comparison with the \textit{Herschel}/PACS images. In the least-squares fitting, a single image is given the same weight as a single photometric data point. 

\subsubsection{Single annulus disk}

In the first instance, we attempt to replicate the outer disk with a single annulus. Whilst the SED is strongly suggestive of a multiple component model, the disk is not well resolved such that a continuous disk should be investigated as a plausible architecture for the HD~76582 system. We account for the mid-infrared excess in this (and the next) model as an unresolved blackbody with a temperature of 150~K. The contribution of this component to the emission in \textit{Herschel} images is small, with flux densities of 6~mJy at 100~$\mu$m and 2~mJy at 160~$\mu$m. 

The system geometry derived from the deconvolution process is adopted for the model architecture, fixing the inclination of the disk, $i$, as 64\degr, and the position angle, $\theta$, at 103\degr. The disk extent is defined by its inner radius, $R$, width, $\Delta R$, and radial surface density profile, $\alpha$. The dust grains are assumed to be compact hard spheres composed of astronomical silicate \citep{Draine2003}, whose size distribution is defined by an exponent $\gamma$ (d$n \propto a^{-\gamma}~$d$a$) and ranges between $a_{\rm min}$ and 1~mm.

We find a good fit to the SED, shown in Fig. \ref{figure:sed_one_part}, the properties of which are summarised in Table \ref{table:one_part_disk} with a broad disk lying from 30 to 230~au with an outwardly increasing surface density profile of $\alpha\!=\!1.5$. Whilst unusual, an outwardly increasing surface density profile is not unheard of amongst debris disks e.g. AU Mic \citep{Macgregor2013} and HD~107146 \citep{Ricci2015}. In order to replicate the lack of strong mid-infrared emission at $\lambda~<~20~\mu$m and the steep submillimetre slope of the SED, a large minimum grain size of 40~$\mu$m and steep particle size distribution exponent of $\gamma~=~5.3$ are preferred in the model. The grain size is around ten times the blow-out radius for the star ($a_{\rm blow}~\sim~3.7~\mu$m) which is a more typical result for later GK-type stars, whilst the steep size distribution implies that the disk emission is dominated by the smallest grains in the disk.

\begin{table}
\caption{Parameters of the single component power law disk model. \label{table:one_part_disk}}
\centering
\tabcolsep 2 pt
\begin{tabular}{lccc}
\hline
Parameter & Range & Distribution & Fit \\
\hline\hline
$R$ [au]        & 10 -- \phantom{0}50 & Linear & 50~$\pm$~5 \\
$\Delta R$ [au] & 10 -- 250 & Linear & 170$^{+40}_{-30}$\\
$\alpha$ [-]    & -5.0 -- \phantom{-}2.0 & Linear & 1.5$^{+0.2}_{-0.5}$ \\
$a_{\rm min}$ [$\mu$m] & 0.5 -- 50 & Logarithmic & 40$~\pm~5$ \\
$a_{\rm max}$ [$\mu$m] & 1000 & Fixed & 1000 \\
$\gamma$ [-]    &  3.0 -- 7.0 & Linear & 5.3$^{+0.2}_{-0.4}$ \\
$M_{\rm dust}$ [$\times10^{-2}~M_{\oplus}$] & n/a & Continuous & 3.5 \\
$\chi^{2}_{\rm red}$ & n/a & n/a & 13.25 \\
\hline
\end{tabular}
\end{table}

\begin{figure*}
\centering
\includegraphics[trim={.5cm .5cm .5cm .5cm},clip,width=\textwidth]{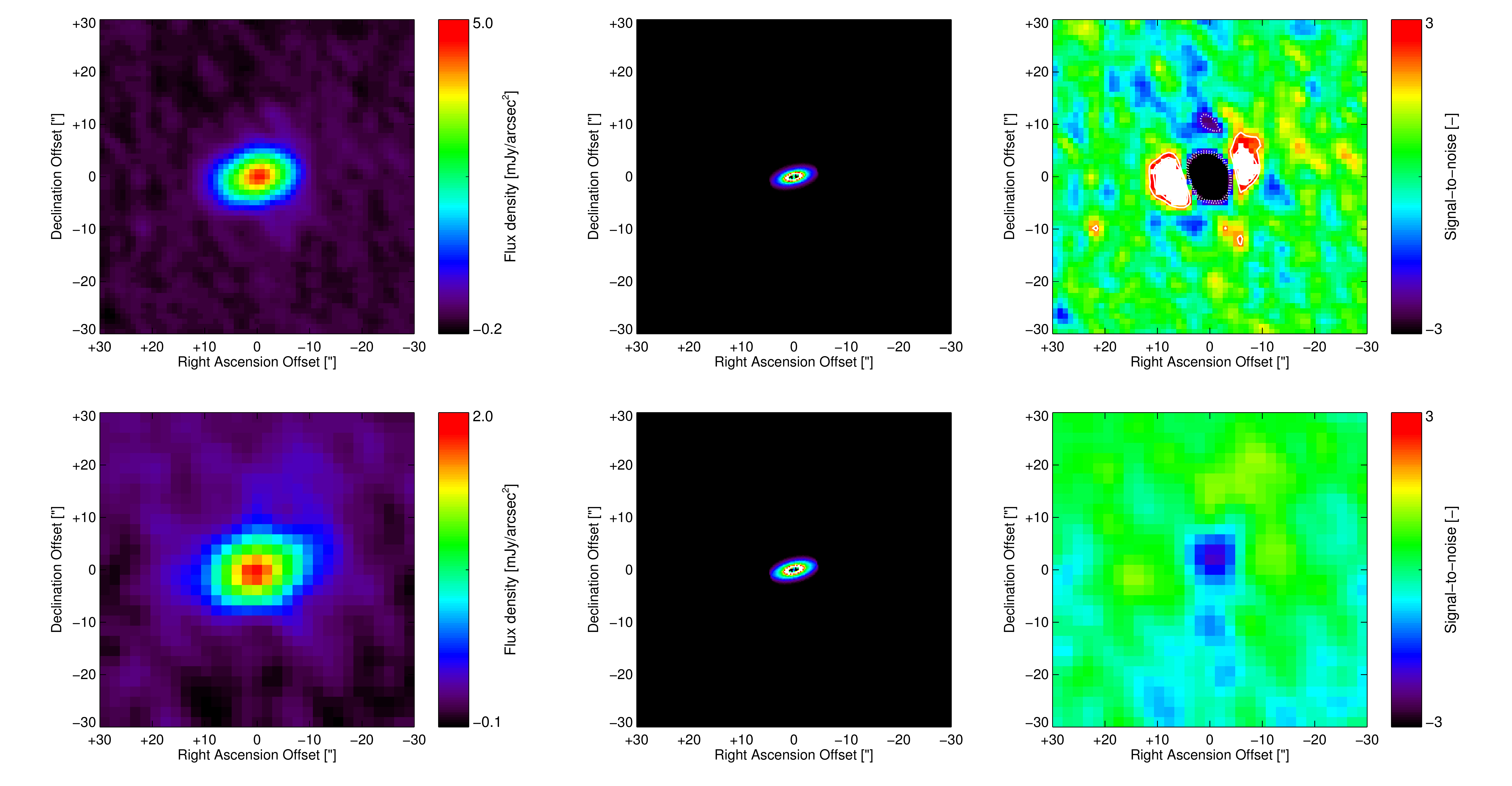}
\caption{Broad disk model fit to the \textit{Herschel}/PACS images at 100~$\mu$m (top) and 160~$\mu$m (bottom). The images show, from left to right, the observations, a high resolution model of the disk, and the residuals after subtraction of the PSF-convolved models from the images. Contours in the residual images are $\pm2\sigma$ and $\pm3\sigma$, with broken contours denoting negative $\sigma$ values. Significant residual flux is clearly visible in the 100~$\mu$m image. \label{figure:image_one_part}}
\end{figure*}

Although the SED is adequately matched by the single component model, the 100~$\mu$m residual image, shown in Fig. \ref{figure:image_one_part} has a large amount of residual flux, at the $\geq~$5-$\sigma$ level, in its outer regions, and suffers from over subtraction in the inner regions. The peak of the residual flux lies at a separation of $\sim~$250~au from the star, and is symmetric either side of the stellar position along the disk major axis. The presence of this residual emission is indicative that a two component model is justified in fitting this disk, which we examine in the next section. 

In Vican et al. (submitted) a single outer component with a radius of 210~au is used to replicate the extended emission in the \textit{Herschel} PACS images. If a model that more heavily weights the disk structure over the SED in the fitting is adopted, we can replicate their findings. In this instance the SED is not well replicated, either having too narrow an excess at far-infrared wavelengths or too shallow a sub-millimetre slope, depending on the dust grain properties adopted. We therefore find that a single component for the outer disk is inadequate to the task of replicating all the available data.

\begin{figure*}
\centering
\includegraphics[trim={.5cm .5cm .5cm .5cm},clip,width=0.8\textwidth]{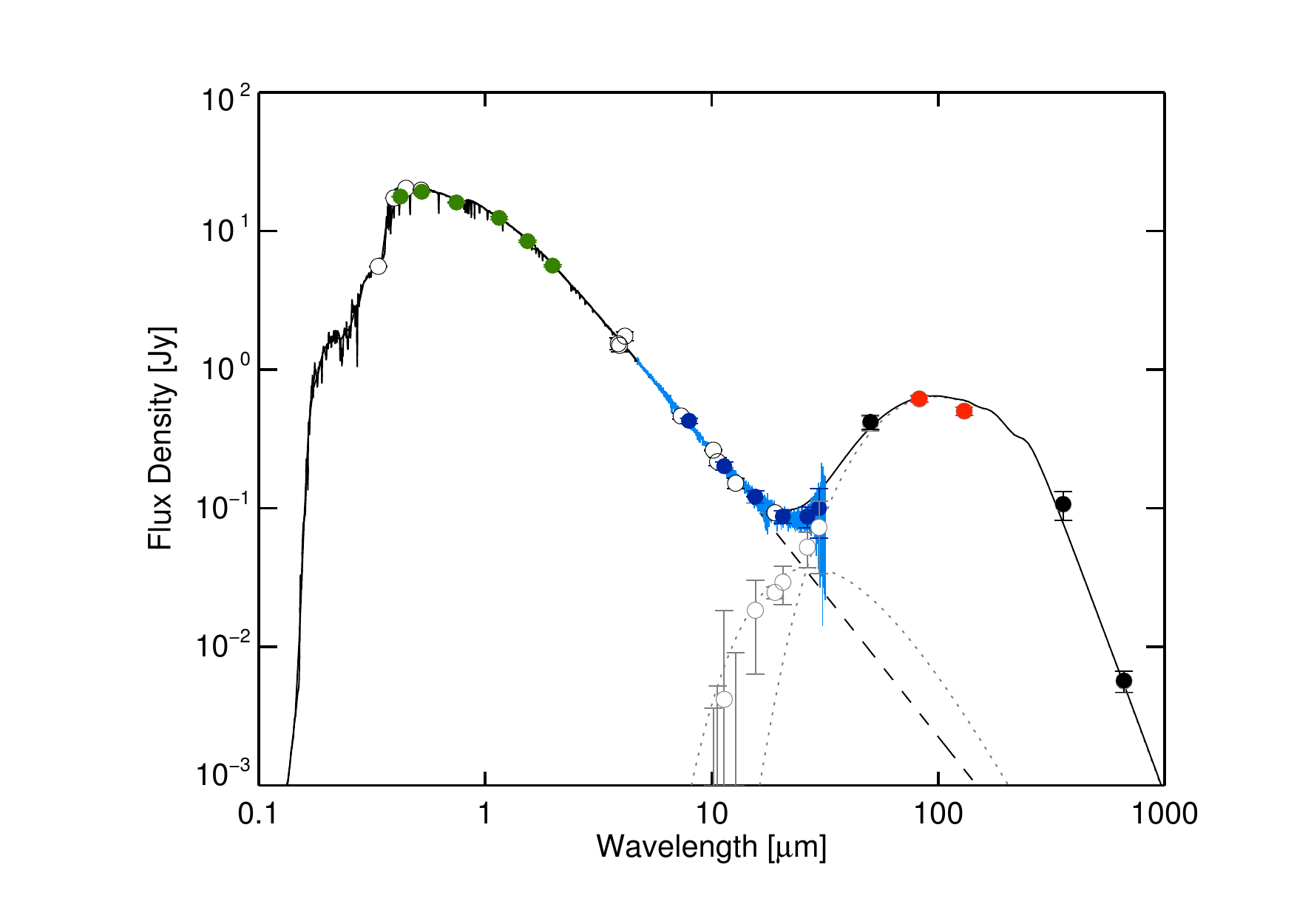}
\caption{Single annulus power law model + blackbody SED fit. Green denotes data used to scale the model stellar photosphere, light and dark blue denote the \textit{Spitzer}/IRS spectrum and photometry respectively, red denotes \textit{Herschel}/PACS photometry, black denotes JCMT/SCUBA-2 photometry, white denotes literature data, and grey denotes photosphere subtracted photometry. The dashed line is the stellar photosphere, the dot-dash line is the disk contribution, and the solid line is the combined star+disk model. \label{figure:sed_one_part}}
\end{figure*}

\subsubsection{Multi-annular disk}

Following the results of the modified blackbody and power law model, we model the outer disk using two annuli (i.e. a total of three belts present in the disk). As with the single annulus model, we use a blackbody component to account for the mid-infrared excess. In this instance the number of model parameters potentially outnumbers the number of data points, due to sparse sampling of the sub-millimetre SED, and the inner component of our disk model is only weakly constrained by the available data due to their low spatial resolution. To aid the modelling process we therefore make several simplifying assumptions regarding the properties of the disk. 

Firstly, we assume both annuli are narrow and co-planar, i.e. they have a width of 10~au and the same inclination and position angle. As per the broad disk model, the inclination, $i$, was assumed to be 64\degr and the position angle, $\theta$, was fixed at 103\degr. Both annuli in the disk are defined by their radial distance (to inner edge of the annulus), $R_{\rm in}$ and $R_{\rm out}$, their widths, $\Delta R_{\rm in}$ and $\Delta R_{\rm out}$. Since the widths of the two components were fixed at 10~au, any investigation of the radial surface brightnesses, $\alpha_{\rm in}$ and $\alpha_{\rm out}$, is unmeaningful, hence these are also fixed as 0. The dust grain properties are defined by the minimum and maximum size of grains in the disk, $a_{\rm min}$ and $a_{\rm max}$, the power-law exponent of the particle size distribution, $\gamma$, and the dust composition \citep[astronomical silicate,][]{Draine2003} with optical properties determined from Mie theory \citep[e.g.][]{Burns1979,BohHuf1983}. We assume that the grain size distribution is the same for both components, and the largest size of dust grains considered in the model is 1~mm. Finally, the slope of the particle size distribution, $\gamma$, is assumed to be the same for both components.

\begin{table}
\caption{Parameters of the two-component power law disk model. \label{table:two_part_disk}}
\centering
\tabcolsep 2 pt
\begin{tabular}{lccccc}
\hline\hline
          &\multicolumn{2}{c}{Range} & Distribution & \multicolumn{2}{c}{Fit}\\
Parameter & Inner & Outer & & Inner & Outer \\
\hline
$R$ [au]                & 10 -- 100     & 100 -- 300   & Linear & 80~$\pm$~5   & 270$^{+30}_{-10}$ \\
$\Delta R$ [au]         & \multicolumn{2}{c}{10} & Fixed & \multicolumn{2}{c}{10} \\
$\alpha$                & \multicolumn{2}{c}{0.0} & Fixed  & \multicolumn{2}{c}{0.0} \\
$a_{\rm min}$ [$\mu$m]  & \multicolumn{2}{c}{0.1 -- 100.0} & Logarithmic & 10$^{+2}_{-3}$ & 36$^{+4}_{-5}$ \\
$a_{\rm max}$ [$\mu$m]  & \multicolumn{2}{c}{1000} & Fixed & \multicolumn{2}{c}{1000} \\
$\gamma$                & \multicolumn{2}{c}{3.0 -- 7.0} & Linear & \multicolumn{2}{c}{5.0$^{+0.5}_{-0.2}$} \\
$M_{\rm dust}$ [$\times10^{-2}~M_{\oplus}$] & \multicolumn{2}{c}{n/a} & Continuous & 0.5 & 7.0 \\
$L_{\rm IR}/L_{\star}$ [$\times10^{-6}$] & \multicolumn{2}{c}{n/a} & Continuous & 145 & 54 \\
$\chi^{2}_{\rm red}$    & \multicolumn{2}{c}{n/a} & n/a & \multicolumn{2}{c}{2.45} \\
\hline
\end{tabular}
\end{table}

\begin{figure*}
\centering
\includegraphics[trim={.5cm .5cm .5cm .5cm},clip,width=\textwidth]{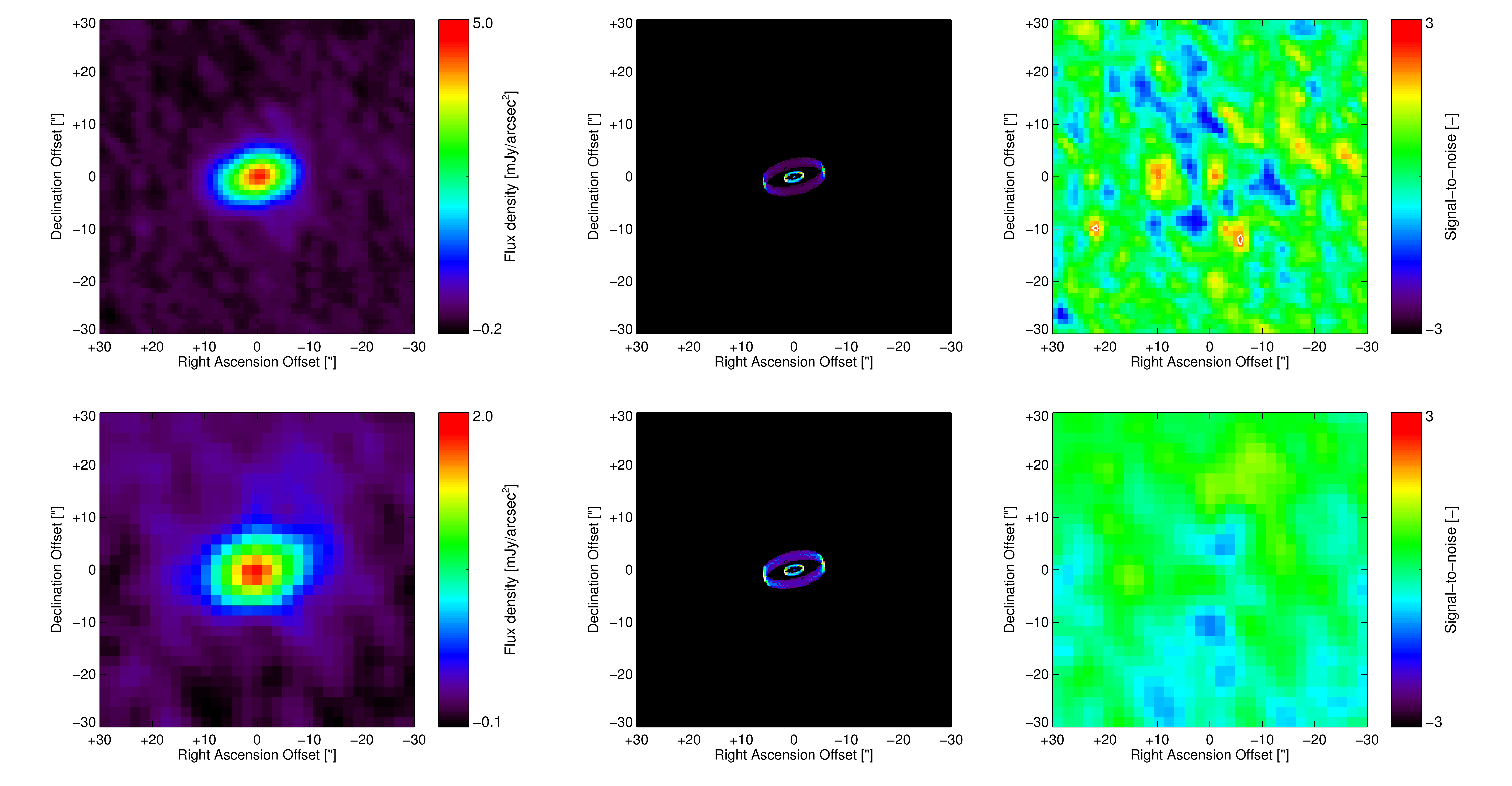}
\caption{Multi-annular disk model fitted to the \textit{Herschel}/PACS images at 100~$\mu$m (top) and 160~$\mu$m (bottom). The panels show, from left to right, the observed disk, a high-resolution two-ring disk model, the disk model convolved with the instrument PSF, and the residuals after subtraction of the convolved model from the observation. Contours in the residual images are $\pm2\sigma$ and $\pm3\sigma$, with broken contours denoting negative $\sigma$ values. \label{figure:image_two_part}}
\end{figure*}

\begin{figure*}
\centering
\includegraphics[trim={.5cm 0cm 0cm .5cm},clip,width=0.8\textwidth]{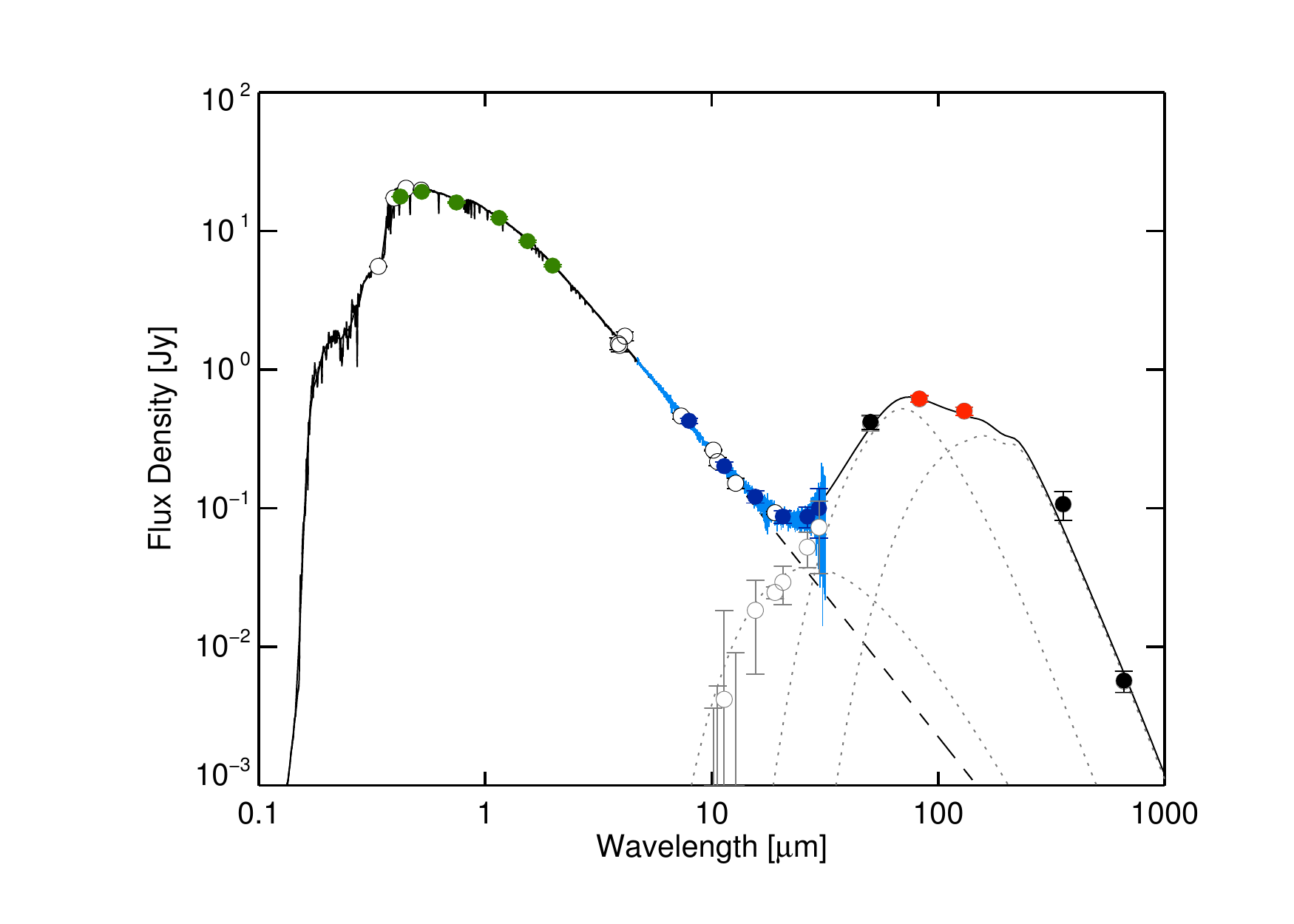}
\caption{Two component power law model + blackbody SED fit. Green denotes data used to scale the model stellar photosphere, light and dark blue denote the \textit{Spitzer}/IRS spectrum and photometry respectively, red denotes \textit{Herschel}/PACS photometry, black denotes JCMT/SCUBA-2 photometry, white denotes literature data, and grey denotes photosphere subtracted photometry. The dashed line is the stellar photosphere, the dot-dash lines are the disk components, and the solid line is the combined star+disk model. \label{figure:sed_two_part}}
\end{figure*}

A two component outer disk model adequately replicates both the SED and extended emision of this disk, as illustrated in Figs \ref{figure:sed_two_part} and \ref{figure:image_two_part}. We find that an outer disk comprised of two narrow annuli at $R_{\rm in}\!=\!$80~au and $R_{\rm out}\!=\!$270~au adequately replicates the disk structure. For the dust properties, a minimum grain size of 10~$\mu$m for the 80~au annulus and 36~$\mu$m for the 270~au annulus are obtained; this result is tied to the need for the cooler (outer) annulus to match the steep sub-millimetre SED without contributing too strongly to the far-infrared part of the SED at 60 and 100~$\mu$m. The particle size distribution is again found to be significantly steeper than for a steady-state case with $\gamma\!=\!5$ but this is also dictated by the sub-millimetre slope. 

\section{Discussion}

Our modelling of HD~76582's disk has shown evidence for three distinct physical components in the disk. Excesses at both mid- and far-infrared wavelengths, requiring disk components at two distinct temperatures, are often interpreted as evidence for physically distinct planetesimals belts in the disk \citep[e.g.][]{Morales2011,Morales2013,Kennedy2014}. As seen above, the disk's observational properties can be adequately replicated by a multi-annular architecture. However, a broad outer disk, over 150~au in extent, cannot replicate both the SED and the extended emission seen in the 100~$\mu$m \textit{Herschel} images, and it would likewise be difficult to justify from a dynamical perspective. We therefore favour the multi-annular interpretation suggesting the outer disk structure has two distinct and physically separate components along with a spatially unresolved inner component. However, images with low spatial resolution provide little constraint to the inner parts of the system; a clear physical separation between the inner and outer system is not conclusively shown in these data. It might well be that the disk is a single contiguous structure around the star, but with several regions with different properties (as we expect a radial variation in the particle size distribution of dust). Multi-annular structures have previously been inferred in this manner for the debris disks around e.g. $\beta$ Leo \citep{Churcher2011}, $\gamma$ Dor \citep{BroekFiene2013}, and HD~23484 \citep[HIP~17439,][]{Ertel2014}. 

The resolved outer disk extent is approximately twice the blackbody radial extent ($A_{\rm disk}/R_{\rm bb}\!=\!2.08$). For a star of $\sim$~10~$L_{\odot}$, the ratio of observed disk radius to blackbody radius, $\Gamma$, is expected to lie around 3.5 \citep[Fig. 10;][]{Booth2013}. Observations by \textit{Herschel} show a range of values for $\Gamma$ up to $\sim$2 \citep{Booth2013,Pawellek2014}. The disk therefore lies closer to the blackbody radius than theoretical expectations, suggesting that the dominant grains responsible for far-infrared emission are larger than the blow out radius ($a_{\rm blow} = 3.7~\mu$m) but not large enough to radiate as blackbodies. This result is consistent with the minimum grain size ($a_{\rm min} = 36~\mu$m) derived in the modelling process.

A typical particle size distribution exponent in the range 3 to 4 is inferred for those debris discs that have been detected at (sub-)millimetre wavelengths \citep{PanSchl2012,Ricci2015}. Such values are consistent with a steady-state collisional cascade and expected values derived from collisional modelling \citep[3.5 to 3.7,][]{Dohnanyi1969,ThebAuger2007,Gaspar2012}. Here, we find an exponent of 5 is required to replicate the SED's sub-millimetre slope. This, in combination with the minimum grain size from the model $a_{\rm min}$ being 36~$\mu$m (for the outer disk), implies the disk (at least in its outer regions) is dominated by moderately-sized dust grains. Such properties are reminiscent of the peculiar ``steep SED'' debris discs detected at far-infrared wavelengths by \cite{Ertel2012}, that have comparable ages ($\sim$300~Myr) to that quoted for HD~76582 (at the lower end), and may therefore represent analogous systems to this one, albeit much fainter ($L_{\rm IR}/L_{\star}~\sim~{\rm few}\times10^{-6}$). For disks around luminous stars, the minimum grain size infered from modelling tends to approach the blowout radius \citep{Booth2013,Pawellek2014}. The mid- and far-infrared excesses preclude the presence of small grains, below $\sim$~10~$\mu$m, contributing to the observed emission as that would increase the disk brightness too much at wavelengths $<$~100~$\mu$m. The minimum grain size infered for the outer parts of this system is thus some 3 to 10 times larger than the blowout radius, 3.7~$\mu$m, derived from radiation pressure arguments \citep{Burns1979}.  

The values of $a_{\rm min}$ in the outer disk derived from the astronomical silicate model are large (10~$\mu$m and 36~$\mu$m). This seems to contradict the high value for $\beta$ found in fitting the modified blackbody model which would imply a large amount of small grains. Comparing the results from the astronomical silicate model, with large grains and a steep size distribution, and the modified blackbody model, with the high beta value, both point towards a disk where its emission is being dominated by the smallest grains. For a more typical disk, a shallower size distribution (in the astronomical silicate model) would broaden the shape of the far-infrared excess and decrease the beta value (for the modified blackbody case). These modelling outcomes can be reconciled by noting that the grains simply need to be inefficient emitters to replicate the sub-mm slope, and that the observed emission is dominated by the smallest grains in the disk (which may not necessarily be small themselves).

A further avenue of investigation is the adoption of a different material for the disk's constituent dust grains. Given the broad parameter space opened by relaxing the condition of a single (pure astronomical silicate) dust composition it is not unreasonable to expect that some combination of materials (and porosity) such as water ice might replicate the disk SED. However, the modelling process is limited such that any multi-material composition found to replicate the SED in this way would exceed the number of available free parameters. That being said, diligence requires us to address that possibility here. We may attempt to fit the SED using dust grains of different properties whilst keeping the constraint of the disk architecture derived from fitting the images. Three alternative materials to astronomical silicate are considered: amorphous water ice \citep{LiGreen1998}, `dirty ice' \citep{Preibisch1993}, and graphite \citep{Jaeger1998}. The optical constants for these materials were taken from the Jena database\footnote{http://www.astro.uni-jena.de/Laboratory/OCDB/}. We found that none of the materials could replicate the SED with an appreciably shallower size distribution or smaller minimum grain size than the results obtained for astronomical silicate grains, and we therefore do not devote any further efforts to presentation of these results. Whilst this approach is non-exhaustive, it demonstrates that the disk properties calculated in this work are not a quirk of adopting standard assumptions regarding the dust grain properties.

Given its unusual properties, we may consider the evolutionary state of the disk; is it actually in a state of flux, or are its properties consistent with a steady-state evolution? To do this, we compare the fractional luminosity of HD~76582's disk to the models presented in \cite{Wyatt2007b}, whereby the maximum brightness, $f_{\rm max}$, for a disk around an A star of a given age can be calculated, under certain assumptions (their Eqns. 14, 20). HD~76582's disk has fractional luminosities of 3.1$\times10^{-5}$ for the warm disk (blackbody), and 1.45$\times10^{-4}$ and 5.4$\times10^{-5}$ for the two annuli in the cool disk. Adopting a separation of $r_{\rm in}~=~20~$au for the blackbody component, and $r_{\rm mid}~=~$80~au and $r_{\rm out}~=~$270~au for the outer two components we can consider these regions separately. Given an age of 0.45~Gyr, $f_{\rm max, in}$ is 1.8$\times10^{-5}$, whereas for an age of 2.13~Gyr, $f_{\rm max, in}$ is 3.9$\times10^{-6}$. For the outer system, at 0.45~Gyr $f_{\rm mid,max}~=~1.1\times10^{-4}$ and $f_{\rm out,max}~=~5.9\times10^{-4}$, whereas at 2.13~Gyr $f_{\rm mid,max}~=~2.5\times10^{-5}$ and $f_{\rm out,max}~=~1.2\times10^{-4}$. Given the large uncertainties on the calculation of $f_{\rm max}$ \citep{Wyatt2007b}, this does not conclusively show that the system is brighter than expected for steady state collisional evolution. Nonetheless, taken together with the steep dust size distribution and large minimum grain size, this does leave open the possibility that HD~76582's disk is not in a quiescent state and is in the process of some dynamical upheaval revealed through its unusual dust emission. The outermost parts of the disk still having a brightness compatible with steady state evolution, would be expected with this scenario as the collisional timescales will be greater there. 

A further issue in examining the state of HD~76582's disk is the evolutionary state of the host star. As an F0~IV sub-giant, the star is undergoing mass loss via a strong stellar wind. The impact of stellar winds on debris disks on the main sequence has been investigated in \cite{Mizusawa2012}, where no difference in excess detection rate between early and late F-type stars was seen, suggesting the stellar winds had little impact on the circumstellar debris. Around sub-giant stars, disks have been examined by e.g. \cite{Bonsor2013,Bonsor2014}. For these post-main sequence disks \citep{BonWya2010} calculated that the disk survives the evolution process but the increased luminosity of the evolved host is more effective at removing smaller grains through radiation pressure. However, the impact of the stellar wind on the disk was negligible over the range of systems studied, as per the main sequence.

In the case of HD~76582, the multi-annular structure of the disk speaks to the presence of massive, unseen bodies in the intervening gaps. As a host star undergoes mass loss on the giant branch, the semi-major axes of companion bodies will expand outwards by a factor of two to three in the adiabatic case \citep{MusVil2012,Mustill2014}. An exoplanet (or large planetesimals) migrating outward in this fashion would stir the planetesimal belts, brightening the disk through increased dust production from collisions. As a subgiant, the amount of mass loss from HD~76582 is small and the impact of this on the orbits of circumstellar bodies would therefore be modest.

To resolve the issue of the disk structure satisfactorially, further high angular resolution images are required. Two obvious approaches are through (sub-)mm interferometry or scattered light imaging at optical/near-infrared wavelengths. However, the HD~76582 system presents a challenge in this regard. In the case of sub-mm follow up, the precipitous drop of the sub-mm excess reduces the integrated flux for the disk to 1.1~mJy at 1300~$\mu$m. Confining this flux to be spread in two annuli within a radial extent of $\sim$~6\arcsec, and assuming a beam FWHM of 1.5\arcsec (to separate the two outer annuli), requires a total of 10 beams to cover the disk area. An observation with s/n $\sim$ 10/beam is feasible with currently available facilities, such as ALMA. The case for scattered light imaging is more difficult, despite the disk being relatively bright, as the conversion between thermal and scattered excess is highly uncertain \citep{Schneider2014}. The large minimum grain size inferred by our model also counts against the feasibility of such a measurement as comparably bright disks with such large grains have been found to be difficult to image, e.g. \cite{Krist2010,Krist2012}.

\section{Conclusions}

We present the first sub-millimetre detection of the debris disk around HD~76582, along with spatially resolved far-infrared images. Measurement of the spectral slope at 450 and 850~$\mu$m reveals a steep fall-off of the disk SED at sub-millimetre wavelengths inconsistent with values observed, and predicted, for typical disks. We combine these new observations with available photometry and images at far-infrared wavelengths to constrain the disk's geometry and extent, and the dust grain properties in a self-consistent model that simultaneously fits all available data. HD~76582's disk has a position angle of 103\degr~and an inclination of 64\degr. The disk modelling favours three physically separate components: an unresolved component at $\sim~$20~au responsible for the mid-infrared excess, and two annuli at 80 and 270~au responsible for the far-infrared excess. The steep size distribution, $\gamma\!=\!5.0$, large minimum grain size, $a_{\rm dust}>$10~$\mu$m, and fractional luminosity of the disk, $L_{\rm IR}/L_{\star}\!=\!2.3\times10^{-4}$, lead us to infer that we do not observe the HD~76582 system in a steady state.

\section*{Acknowledgments}

We thank the referee for improving the paper with their helpful comments. JPM is supported by a UNSW Vice-Chancellor's Postdoctoral Research Fellowship. MB acknowledges support from FONDECYT Postdoctral Fellowships, project no. 3140479, and the Millennium Science Initiative (Chilean Ministry of Economy), through grant Nucleus P10-022-F. This research has made use of the SIMBAD database, operated at CDS, Strasbourg, France. This research has made use of NASA's Astrophysics Data System. This paper is based on data obtained by the JCMT Debris Disc Legacy Survey, Program ID MJLSD02. The James Clerk Maxwell Telescope has historically been operated by the Joint Astronomy Centre on behalf of the Science and Technology Facilities Council of the United Kingdom, the National Research Council of Canada and the Netherlands Organisation for Scientific Research. Additional funds for the construction of SCUBA-2 were provided by the Canada Foundation for Innovation.

\bibliographystyle{mnras}

\bsp

\label{lastpage}

\end{document}